\newcommand\aj{{AJ\,}}%
\newcommand\apj{{ApJ\,}}%
\newcommand\apjs{{ApJS\,}}%
\newcommand\aap{{A\&A\,}}%
\newcommand\mnras{{MNRAS\,}}%
\documentclass[graybox, natbib, footinfo]{svmult}

\usepackage{mathptmx}       % selects Times Roman as basic font
\usepackage{helvet}         % selects Helvetica as sans-serif font
\usepackage{courier}        % selects Courier as typewriter font
\usepackage{type1cm}        % activate if the above 3 fonts are
\usepackage{makeidx}         % allows index generation
\usepackage{graphicx}        % standard LaTeX graphics tool
\usepackage{multicol}        % used for the two-column index
\usepackage[bottom]{footmisc}% places footnotes at page bottom

\makeindex             % used for the subject index
\begin{document}
\title*{Preliminary Evaluation of the {\it Kepler} Input Catalog Extinction Model Using Stellar Temperatures}
\titlerunning{Evaluation of the KIC Extinction Model}
% Use \titlerunning{Short Title} for an abbreviated version of
% your contribution title if the original one is too long
\author{Gail Zasowski, Deokkeun An, and Marc Pinsonneault}
% Use \authorrunning{Short Title} for an abbreviated version of
% your contribution title if the original one is too long
\institute{Gail Zasowski \at The Ohio State University, Columbus, OH, USA, \\ Johns Hopkins University, Baltimore, MD \email{gail.zasowski@gmail.com}
\and Deokkeun An \at Ewha Womans University, Seoul, South Korea, \email{deokkeun.an@gmail.com}
\and Marc Pinsonneault \at The Ohio State University, Columbus, OH, USA, \email{pinsonneault.1@osu.edu}}
%
% Use the package "url.sty" to avoid
% problems with special characters
% used in your e-mail or web address
%
\maketitle

%\vspace{-90pt}
\abstract{
%Each chapter should be preceded by an abstract (10--15 lines long) that summarizes the content. The abstract will appear \textit{online} at \url{www.SpringerLink.com} and be available with unrestricted access. This allows unregistered users to read the abstract as a teaser for the complete chapter. As a general rule the abstracts will not appear in the printed version of your book unless it is the style of your particular book or that of the series to which your book belongs.
The {\it Kepler} Input Catalog (KIC) provides reddening estimates for its stars, based on the assumption of a simple
exponential dusty screen.
This project focuses on evaluating and improving these reddening estimates for the KIC's giant stars, for which extinction
is a much more significant concern than for the nearby dwarf stars.  We aim to improve the calibration (and thus consistency) amongst
various photometric and spectroscopic temperatures of stars in the {\it Kepler} field by removing systematics due to incorrect extinction assumptions.
The revised extinction estimates may then be used to derive improved stellar and planetary properties.
%As the ISM is not only something to be removed from data but something of scientific interest in its own right, 
We plan to eventually use the large number of KIC stars as probes into the structure and properties of the Galactic ISM.
}

\section{Introduction} \label{sec:intro}
%Instead of simply listing headings of different levels we recommend to
%let every heading be followed by at least a short passage of text.
%Further on please use the \LaTeX\ automatism for all your
%cross-references and citations. And please note that the first line of
%text that follows a heading is not indented, whereas the first lines of
%all subsequent paragraphs are.
The dusty interstellar medium (ISM) of the Milky Way (MW) permeates the entire galaxy, 
with non-zero reddening observed even at the Galactic poles. % ([[cite?]]).  
Even though the effects of extinction at infrared (IR) wavelengths are smaller than those at optical wavelengths, 
both optical and near-IR images of the MW demonstrate that interstellar 
dust is strongly concentrated in the midplane but not exclusively confined there.
And the {\it Kepler} field, lying $5^\circ-20^\circ$ off the midplane, is hardly free of its effects.
One of the major problems with unaccounted-for extinction is that it changes properties inferred from photometry (e.g., luminosity, distance, temperature)
in a {\it systematic} way --- stars always look cooler and fainter, never the other way around.

Of course, much effort has been made to address this issue, 
leading to a large number of extinction maps and models derived using a variety of methods and ISM tracers.
To name just a few examples, \cite{Drimmel_2003_3dMWextinction} and \cite{Drimmel_2001_3dMWextinction} have built an analytic model of the MW's ISM, 
complete with a smooth disk and superimposed dusty spiral arms.
The commonly adopted all-sky reddening maps by \cite{Schlegel_1998_dustmap} use primarily the dust emission at 100~$\mu$m to trace extinction, 
along with some assumptions about the dust homogeneity that make the maps tricky to apply within 20$^\circ$ or so of the midplane.
\cite{Marshall_2006_3dMWextinction} compared NIR color-magnitude diagrams to those predicted by the Besan\c{c}on MW stellar populations model
and published extinction maps spanning the MW's midplane.
Very recently, \cite{Gonzalez_2012_bulgeextmap} published maps of the Galactic bulge, using red clump stars from the VVV to trace extinction
on arcmin scales.  

In comparison, the {\it Kepler} Input Catalog (KIC) assumed the extinction model of a smooth, vertically-exponential dust disk, 
with a scaleheight $H$ of 150 pc and an extinction density $\kappa_0$ normalized to 1 mag of $V$-band extinction per kpc at $b=0^\circ$:
\begin{equation}
E(B-V)_{\rm KIC} = 0.367 \kappa_0 \int_0^d e^{-\frac{s}{H}\sin{b}} {\rm d}s, %\nonumber
\end{equation}
\citep[adapted from][]{Brown_2011_KIC}.  Thus the reddening can be parameterized as a function of distance $d$ and latitude $b$ alone,
with typical values for the KIC stars of $E(B-V) \sim 0.05-0.18$ mag.
However, the KIC stellar parameters were derived simultaneously along with the reddening,
meaning that the final $E(B-V)$ values are implicitly tied to all of the parameters, including the effective temperature ($T_{\rm eff}$) and the metallicity.
These correlations mean that the catalog reddening values have strong implications for other properties of interest further down the pipeline,
such as planetary radii or the stellar distances derived using the KIC parameters in other methods,
which are important for any Galactic structure work or study of metallicity or age gradients (to name but a few possibilities).

\section{Comparison of Photometric and Spectroscopic Temperatures} \label{sec:comp_teffs}
\begin{figure} %[h]
\begin{center}
%\sidecaption %[t]
%% Use the relevant command for your figure-insertion program
%% to insert the figure file.
%% For example, with the graphicx style use
  \includegraphics[trim=5in 0in 0.2in 0in, clip, width=0.35\textwidth]{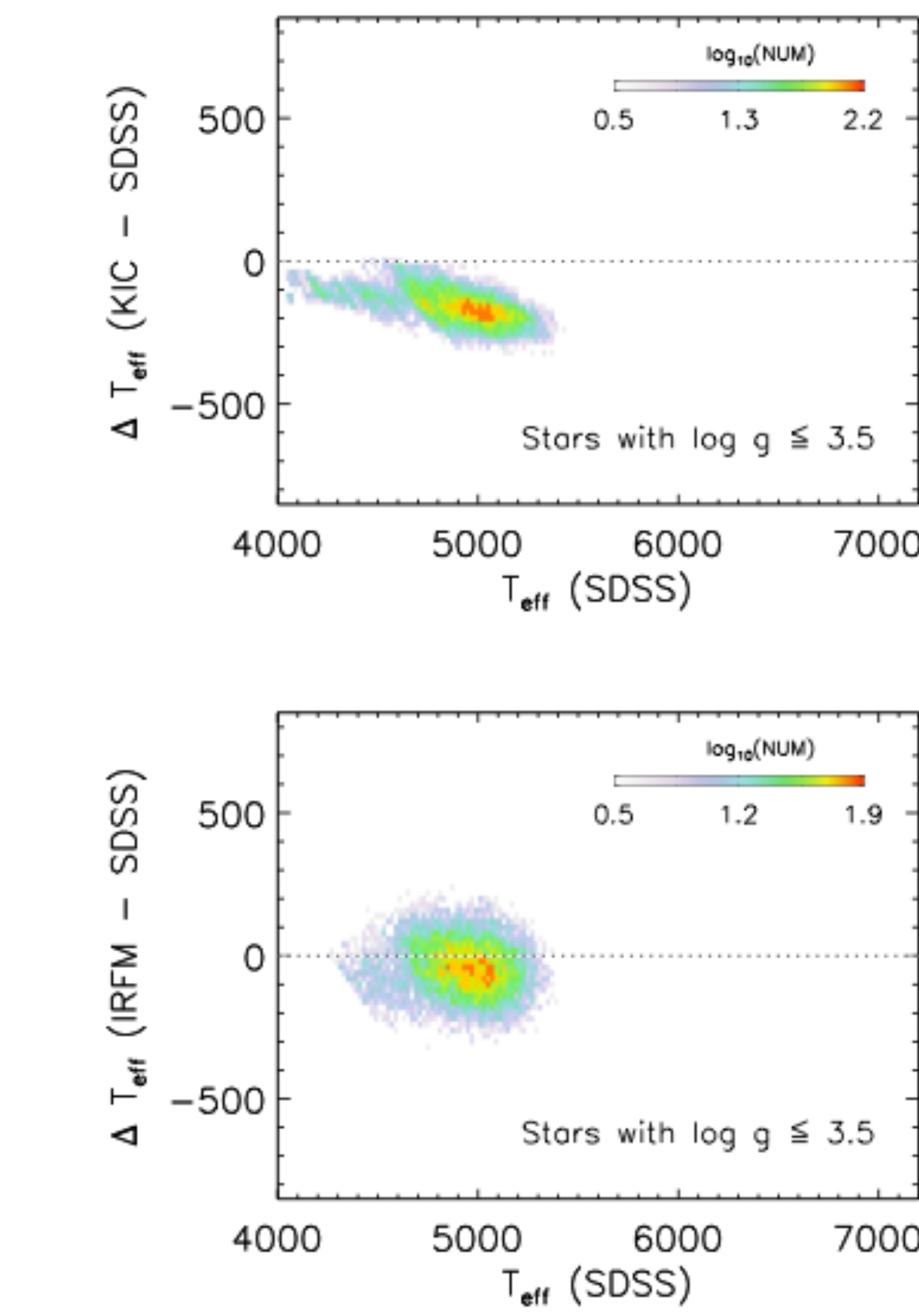}
  \includegraphics[trim=5.5in 1.1in 1.25in 4.5in,clip,width=0.35\textwidth]{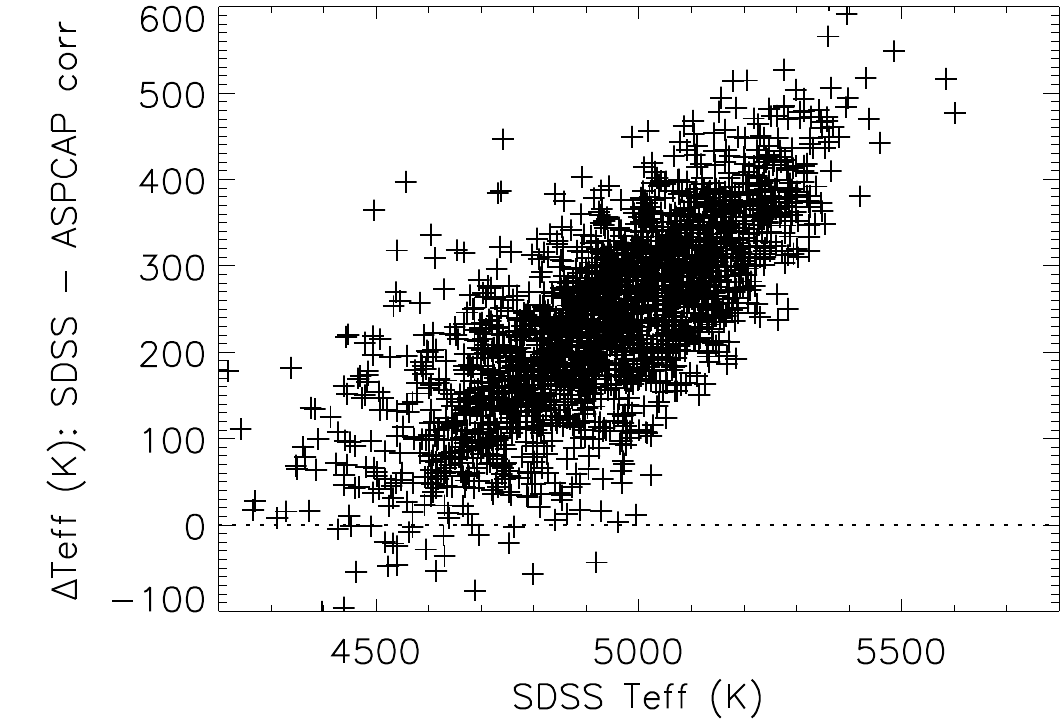}
%%
%% If no graphics program available, insert a blank space i.e. use
%%\picplace{5cm}{2cm} % Give the correct figure height and width in cm
%%
\caption{{\it Left:} Comparison between the SDSS {\it griz} $T_{\rm eff}$ values and (top) the KIC $T_{\rm eff}$ and
    (bottom) IRFM $T_{\rm eff}$ for giant stars.  \citep[Figure 18 of ref][]{Pinsonneault_2013_temprevisionserr}.
    {\it Right:} Difference between the SDSS {\it griz} $T_{\rm eff}$ values and those from APOGEE/ASPCAP, as a function of SDSS $T_{\rm eff}$, 
    for the $\sim$2000 giants in the APOKASC sample.}
\label{fig:sdss-irfm}       % Give a unique label
\end{center}
\end{figure}

In the top left panel of Figure~\ref{fig:sdss-irfm}, we show the comparison between two photometric temperature methods: the original KIC $T_{\rm eff}$ values and those
derived from SDSS {\it griz} colors by \cite{Pinsonneault_2012_temprevisions}.  An offset of $\sim$100--200~K demonstrates that adopting
the KIC reddening values, along with internally consistent color-$T_{\rm eff}$ relations, would give rise to systematically hotter temperatures.
The bottom panel of Figure~\ref{fig:sdss-irfm} shows that the {\it griz} temperatures (corrected for giant stars)
are far more consistent with those derived with the 
InfraRed Flux Method \citep[IRFM;][]{Casagrande_2010_irfm},
which arguably represent a more fundamental measurement of effective temperature as it is defined.  
%For the temperatures calculated using methods originally calibrated for dwarf stars ({\it griz} and IRFM), 
%corrections have been applied to make them appropriate for giant stars. 

At least two studies have targeted {\it Kepler} stars for spectroscopic followup and evaluation of the KIC stellar parameters.
\cite{Thygesen_2012_keplerfollowup} and \cite{Molenda-Zakowicz_2013_keplerfollowup} observed a total of $\sim$200 stars, 
and both found an offset of up to a couple hundred K \citep[which was strongly $T_{\rm eff}$-dependent in][]{Thygesen_2012_keplerfollowup}.

A much larger spectroscopic comparison sample is provided by the APOKASC program, a collaboration between APOGEE,
a high-resolution, $H$-band spectroscopic survey of MW red giant stars \citep{Majewski_2012_apogee}, 
and KASC, the core working group of {\it Kepler} asteroseismic science.
The APOKASC team is combining asteroseismic and spectroscopic measurements and their respective derived parameters for $\sim$10$^4$ stars
\citep[see Section 8.3 of ][]{Zasowski_2013_apogeetargeting}.  
The APOKASC catalog at the time of this proceeding contains $\sim$2000 giant stars, an increase of nearly an order of magnitude over the previous sample.
Figure~\ref{fig:sdss-irfm} also shows the comparison between the photometric {\it griz} $T_{\rm eff}$ and the spectroscopic APOGEE $T_{\rm eff}$, 
where the latter have been ``corrected'' to a temperature scale calibrated to a number of well-studied star clusters.
Despite this correction, a strongly temperature-dependent offset is clearly present, in addition to a $\pm$100~K scatter.
Possible reasons for the persistent discrepancy include systematic offsets in the APOGEE pipeline and/or potential errors
in the theoretical corrections applied to the dwarf-derived {\it griz}-$T_{\rm eff}$ relations to make them suitable for giants.
%(Clearly, additional work needs to be done on the APOGEE pipeline to understand this $T_{\rm eff}$ dependence and offset.)

%\begin{figure}[!h]
%\begin{center}
%%\sidecaption %[t]
%%\vspace{-36pt}
%%  \includegraphics[angle=90,trim=3.3in 0.25in 0in 4.2in,clip,width=0.5\textwidth]{apokasc_spec-vs-phot-teffs}
%  \includegraphics[angle=180,trim=1in 0.25in 5.5in 4.25in,clip,width=0.5\textwidth]{apokasc_spec-vs-phot-teffs.pdf}
%  \caption{Difference between the IRFM $T_{\rm eff}$ values and those from APOGEE/ASPCAP, as a function of IRFM $T_{\rm eff}$, 
%    for the $\sim$2000 giants in the APOKASC sample.}
%  \label{fig:irfm-vs-aspcap}
%\end{center}
%\end{figure}

\subsection{Impact of Extinction?} \label{sec:impact_ext}
%With potential extinction systematics at the forefront of our thoughts, we asked what reddening values would be needed to bring
%the IRFM and APOGEE temperatures in line with each other.  It turns out that good agreement could be achieved by assuming $E(B-V) \sim 0$,
%which is unrealistic.  This exercise serves as an example of how the various factors and uncertainties impacting the stellar parameters
%(here, focusing on $T_{\rm eff}$) can work together, sometimes enhancing each other and sometimes hiding the problems.

\begin{figure} %[!ht]
\begin{center}
%\sidecaption[t]
  \includegraphics[trim=0in 4.5in 0in 0.4in, clip, width=0.4\textwidth]{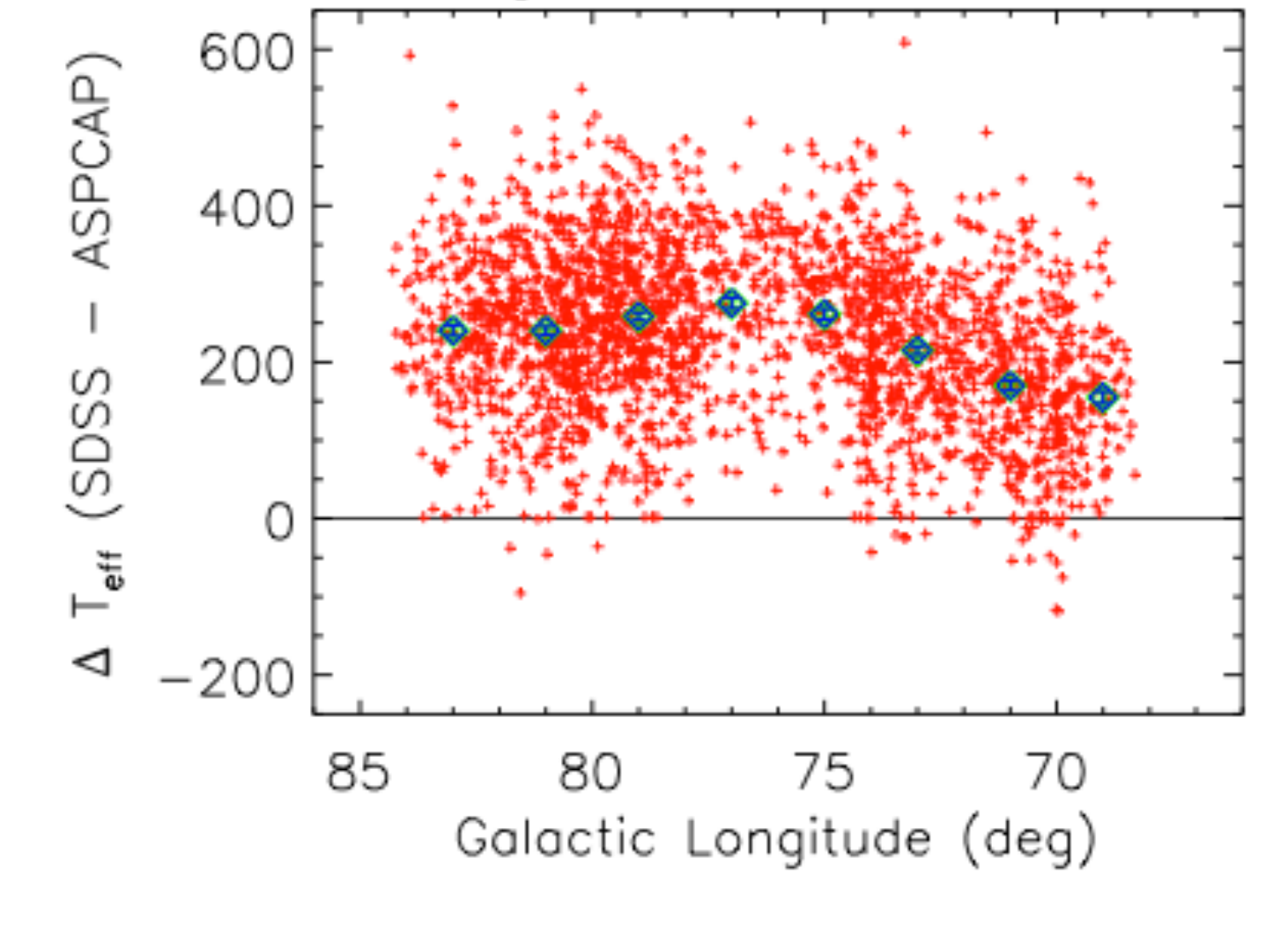}
  \includegraphics[trim=0in 0in 0in 5in, clip, width=0.4\textwidth]{apokasc_sdss-vs-aspcap-vs-position.pdf}
  \caption{Difference in {\it griz} $T_{\rm eff}$ and APOGEE/ASPCAP $T_{\rm eff}$ as a function of (left) Galactic longitude
    and (right) Galactic latitude, assuming the original KIC reddening.}
  \label{fig:sdss-aspcap-position}
\end{center}
\end{figure}

Figure~\ref{fig:sdss-aspcap-position} shows the difference between the {\it griz} temperatures 
(which assume a reddening value) and the APOGEE temperatures (which do not),
as functions of Galactic longitude (left) and latitude (right).
There is no monotonic trend with longitude, though there are hints of coherent structure, but there is a systematic trend with latitude.
As any observed $T_{\rm eff}$ offsets should be independent of spatial position, the behavior seen here suggests a problem with the extinction model,
which is smooth in longitude but dependent on latitude.
In order to test the form and scale of the KIC extinction model, we require a large number of KIC stellar extinction estimates that do not
depend on any global Galactic model (which would unavoidably include its own assumptions about the underlying dust structure).

%\section{RJCE Extinctions} \label{sec:rjce}
\section{Revision of the KIC Extinctions} \label{sec:rev_kic}

\subsection{RJCE Extinctions} \label{sec:rjce}
For independent extinction estimates, we turn to the Rayleigh-Jeans Color Excess method \citep[RJCE;][]{Majewski_2011_rjce},
which provides stellar foreground extinction estimates based on the star's near- and mid-IR photometry.
Like other color excess methods, %(e.g., Lada et al.\,1994; Lombardi \& Alves 2001), 
the basis of RJCE is the assumption that all stars in a given sample have a common
intrinsic color, such that any observed excess in that color may be attributed to line-of-sight reddening.
Where RJCE improves upon earlier color excess approaches is in the combination of NIR+MIR photometry,
which sample the Rayleigh-Jeans (RJ) tail of a star's spectral energy distribution (SED).
The slope of the RJ tail is almost entirely independent of stellar temperature or metallicity,
which means that colors comprising filters measuring this part of the SED are truly homogeneous across
a very large fraction of the MW's stellar populations.  In contrast, color excess methods relying on
more temperature-dependent optical or NIR-only colors must know a priori, or assume, the spectral type of the stars under consideration.
See \cite{Majewski_2011_rjce} for an expanded description and example applications of the RJCE approach.

Most relevant for the current analysis is that the RJCE extinction values are free from assumptions about
the spatial distribution of the MW's dust content and thus provide a good comparison to the KIC values,
which are strongly dependent on the KIC's assumed dust disk model.
Also relevant are some caveats on RJCE's applicability.  Due to [Fe/H]-dependent spectral features,
very metal-poor stars (${\rm [Fe/H]} < -1.5$) appear to have intrinsic colors deviating from those assumed in RJCE,
such that their extinction values are systematically overestimated.
Fortunately, such stars are relatively rare in the {\it Kepler} field.
The uncertainties in the RJCE extinction values arise from the photometric uncertainties as well as uncertainty in the adopted extinction law;
typical values for the KIC sample are $\sigma(A[K_s]) < 0.05$ mag.

\subsection{Extinction Map Comparison} \label{sec:approach}
For this initial assessment of the KIC extinction, we used stars meeting the following criteria: 
$A(K_s)_{\rm RJCE} \ge 0$, $\sigma(A[K_s])_{\rm RJCE} \le 0.05$, 2MASS nearest neighbor distance ({\it prox}) of $\ge$10$^{\prime\prime}$, 
$4000 \le (T_{\rm eff})_{\rm KIC} \le 5000$~K, and $(\log{g})_{\rm KIC} \le 3.7$.
These requirements produced a sample of $\sim$13\,500 red giant stars with well-measured RJCE extinction values.

This sample includes stars as faint as $H \sim 15$, while the spectroscopic APOKASC sample is limited to stars with $H \le 11$.
Because we want to ensure that any corrections calculated are applicable to the stars being corrected, 
we performed the analysis below both with and without an $H \le 11$ limit on the sample.  
We found that this limit actually had little-to-no impact on the outcome, which is a result of the fact that stars
in our sample with $H \le 11$ have a very similar extinction distribution (in RJCE and the KIC) as those with $11 < H < 15$.
Since the fainter giants comprise more distant stars, as well as intrinsically fainter ones, this result may be somewhat surprising,
given the assumption of a strongly distance-dependent extinction distribution.  What this suggests is that a foreground extinction screen
model may be appropriate for this particular sample (certainly not generally applicable to MW stellar populations!), 
which provides further justification for the approach adopted below (in which we do not apply the $H \le 11$ limit).  
Of course, the KIC extinction values {\it are} distance dependent,
and there is a slightly larger discrepancy between the $E(B-V)$ distributions of the $H \le 11$ and $11 < H < 15$ subsamples 
than between the RJCE reddening distributions of the same subsamples, but the discrepancy is still not large enough to significantly affect the outcome below.

\begin{figure}[!h]
\begin{center}
  \includegraphics[trim=1in 1in 1in 1in, clip, width=2.2in]{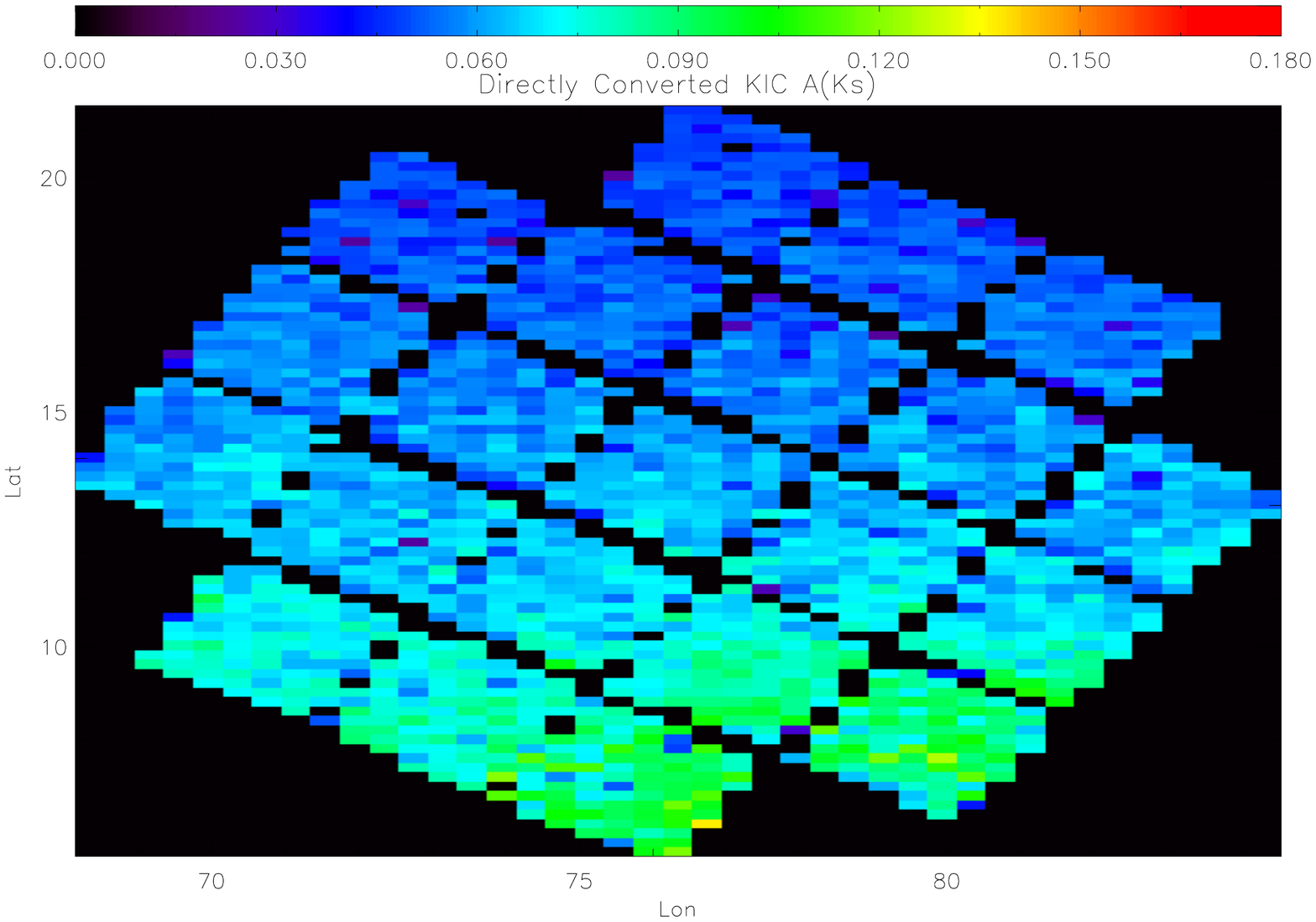}
  \includegraphics[trim=1in 1in 1in 1in, clip, width=2.2in]{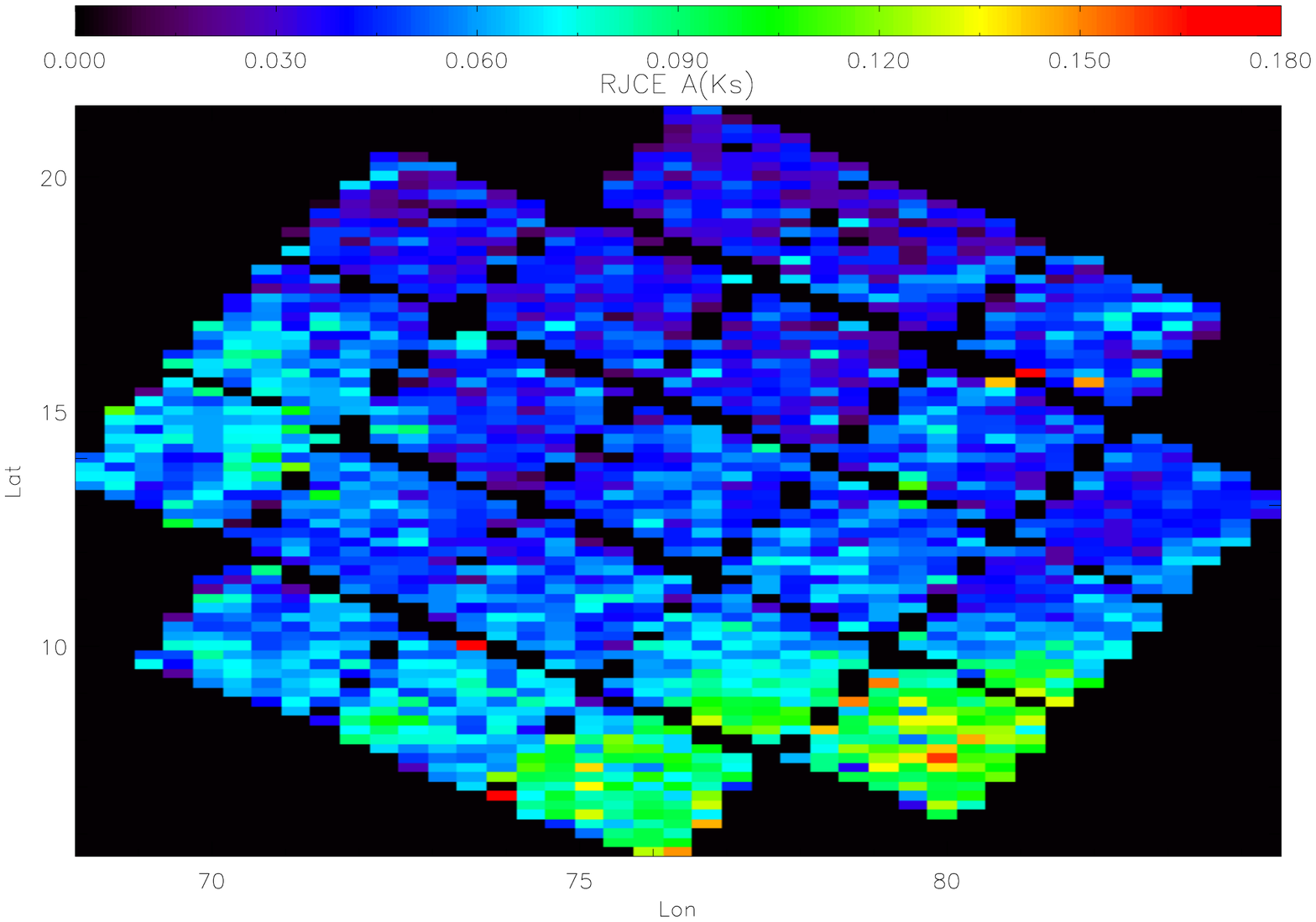}
  \caption{{\it Left:} Median map of KIC $A(K_s)$ values, assuming $A(K_s) = 0.388 \times E(B-V)$.  {\it Right:} Median map of RJCE $A(K_s)$ values.}
  \label{fig:maps}
\end{center}
\end{figure}

We created two extinction maps spanning the entire {\it Kepler} field, taking as the value of each 0.4$^\circ$$\times$0.2$^\circ$ 
pixel the median KIC $E(B-V)$ or RJCE $A(K_s)$ extinction of the stars in that pixel (Figure~\ref{fig:maps}, 
where we have converted the KIC $E(B-V)$ to $A(K_s)$ for comparison purposes).  
This pixel size was chosen to ensure $\ge$10 stars per bin, and the asymmetry reflects the fact that in both maps, 
the variations in $b$ have a shorter scale than those in $l$.
We chose to parameterize this first ``correction'' with a zero-point offset, an extinction-dependent scale factor, and a $b$-dependent scale factor.

\subsection{Results} \label{sec:results}
\begin{figure}[!h]
\begin{center}
  \includegraphics[trim=1in 1in 1in 1in, clip, width=2.2in]{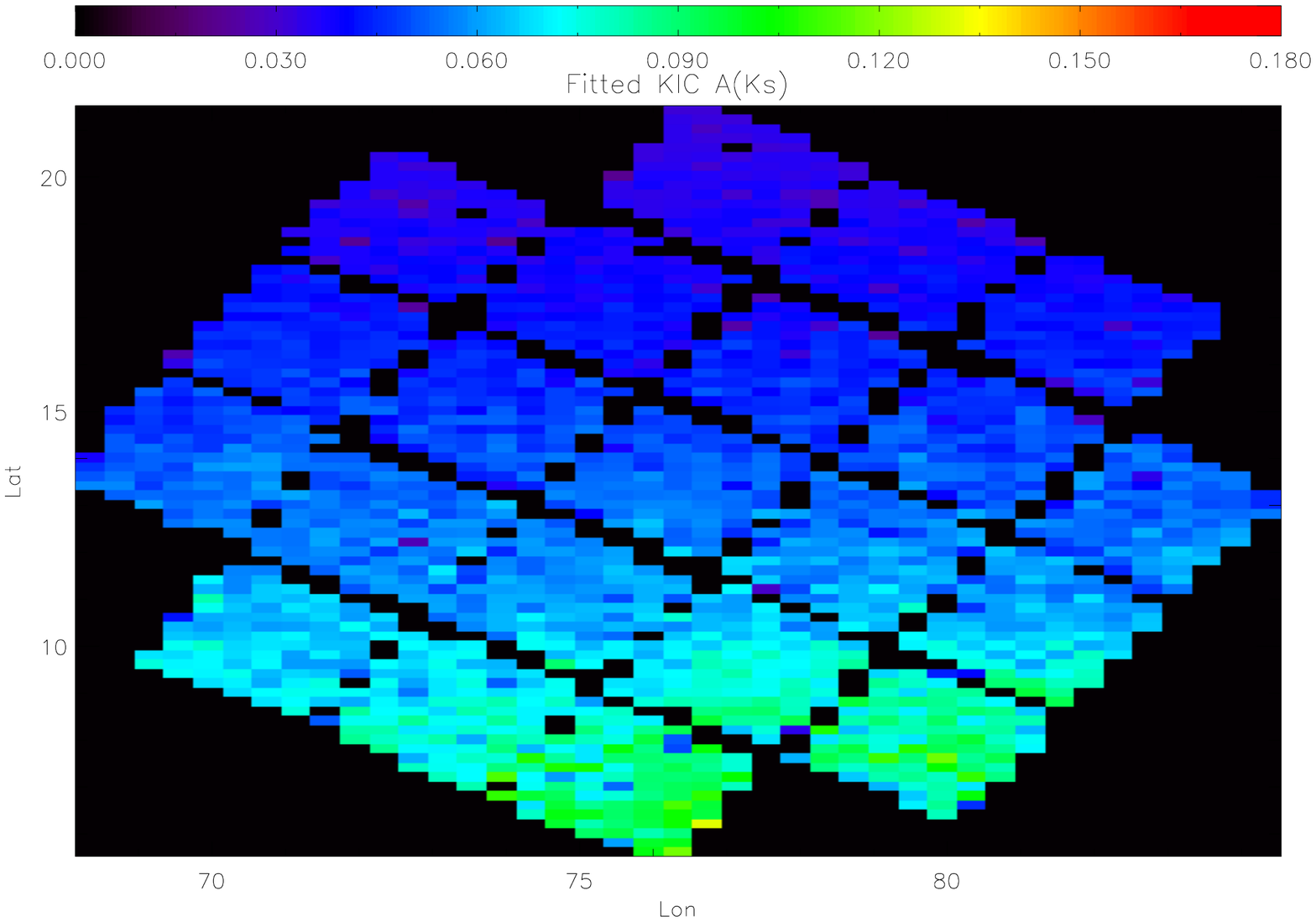}
  \includegraphics[trim=1in 1in 1in 1in, clip, width=2.2in]{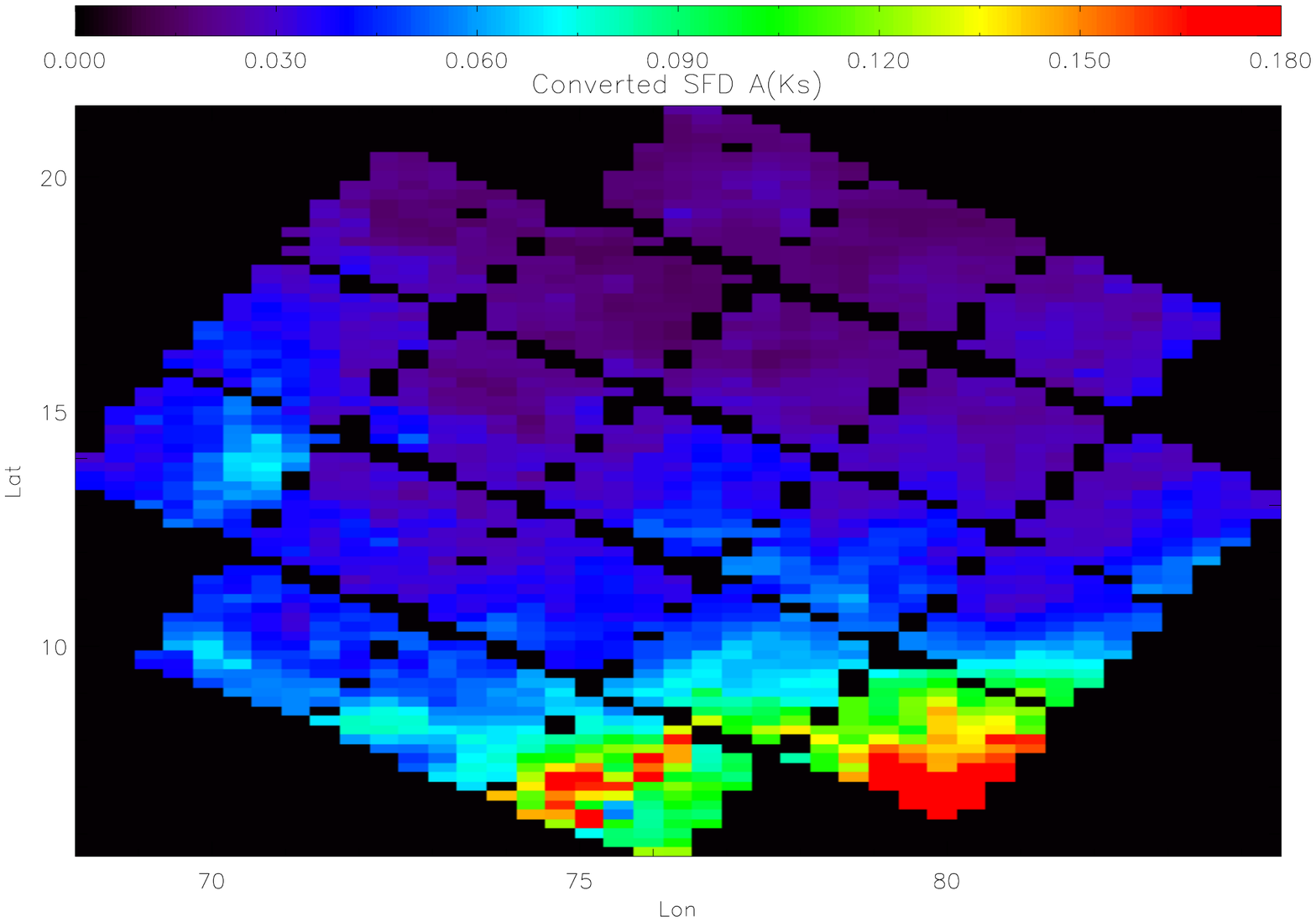}
  \caption{{\it Left:} Median map of revised KIC $A(K_s)$ values.  {\it Right:} Median map of \cite{Schlegel_1998_dustmap} $A(K_s)$ values.}
  \label{fig:maps2}
\end{center}
\end{figure}

The fitting of the KIC map to the RJCE one yields: $E(B-V) = E(B-V)_{\rm KIC} \times (1.29 - 1.75\,\sin{b}) + 0.02$.
The median extinction map using these fitted KIC reddening values is shown in the left panel of Figure~\ref{fig:maps2}.
The impact of this correction scaling is to reduce the effective scaleheight of the KIC model's dust layer,
whose original value was even noted in \cite{Brown_2011_KIC} to be larger than the literature suggested,
and to bring the reddening values more in line with those calculated from the independent RJCE method 
(i.e., compare Figure~\ref{fig:maps} [right] with Figure~\ref{fig:maps2} [left]).
As a further sanity check, we compare the revised KIC map with the \cite{Schlegel_1998_dustmap} dust map (Figure~\ref{fig:maps2}, right), 
and we see much better agreement with the revised map than with the original KIC one, in terms of latitude dependence.

\begin{figure} [!h]
\begin{center}
  \includegraphics[width=2in]{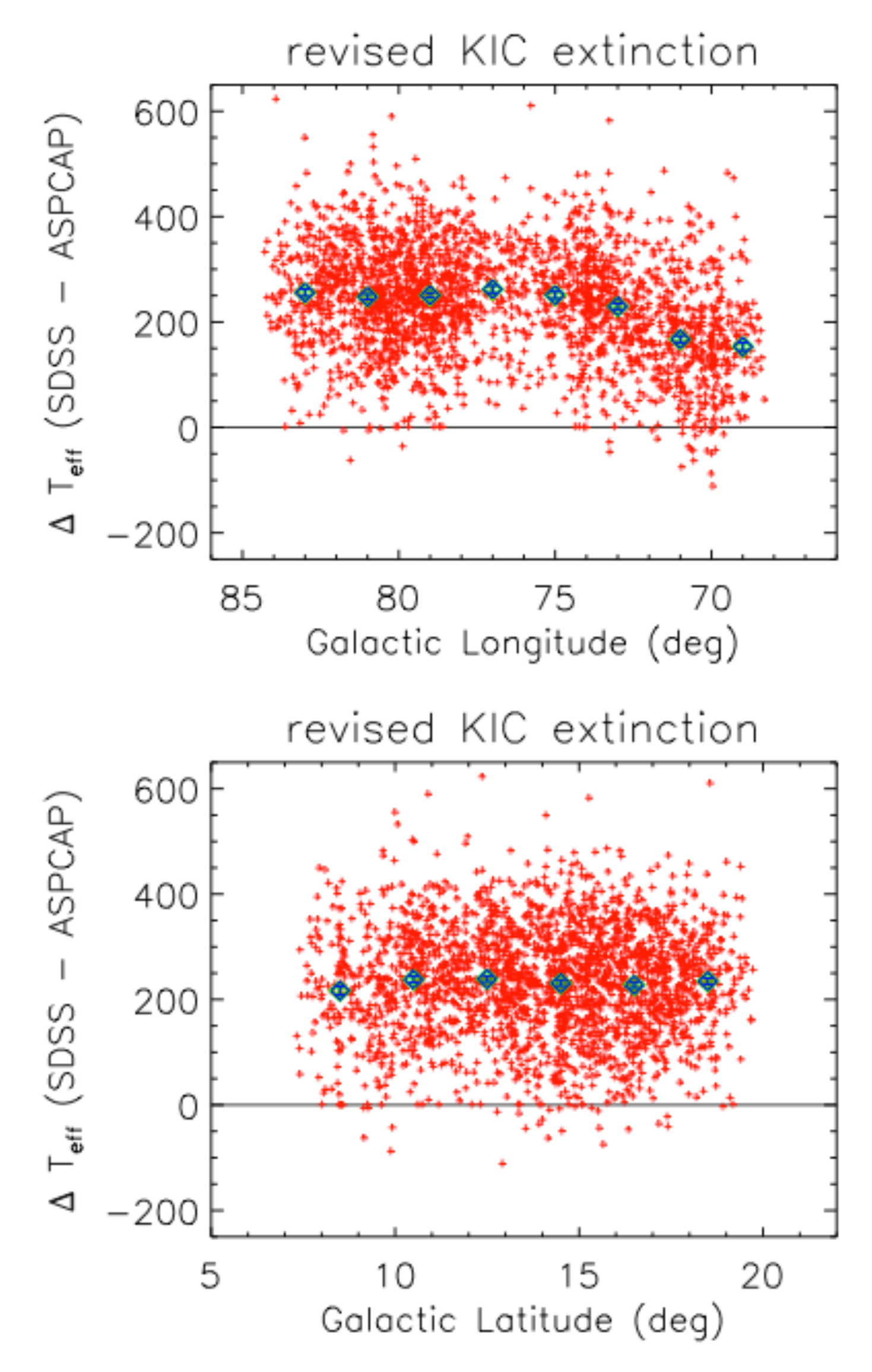}
  \includegraphics[width=2in]{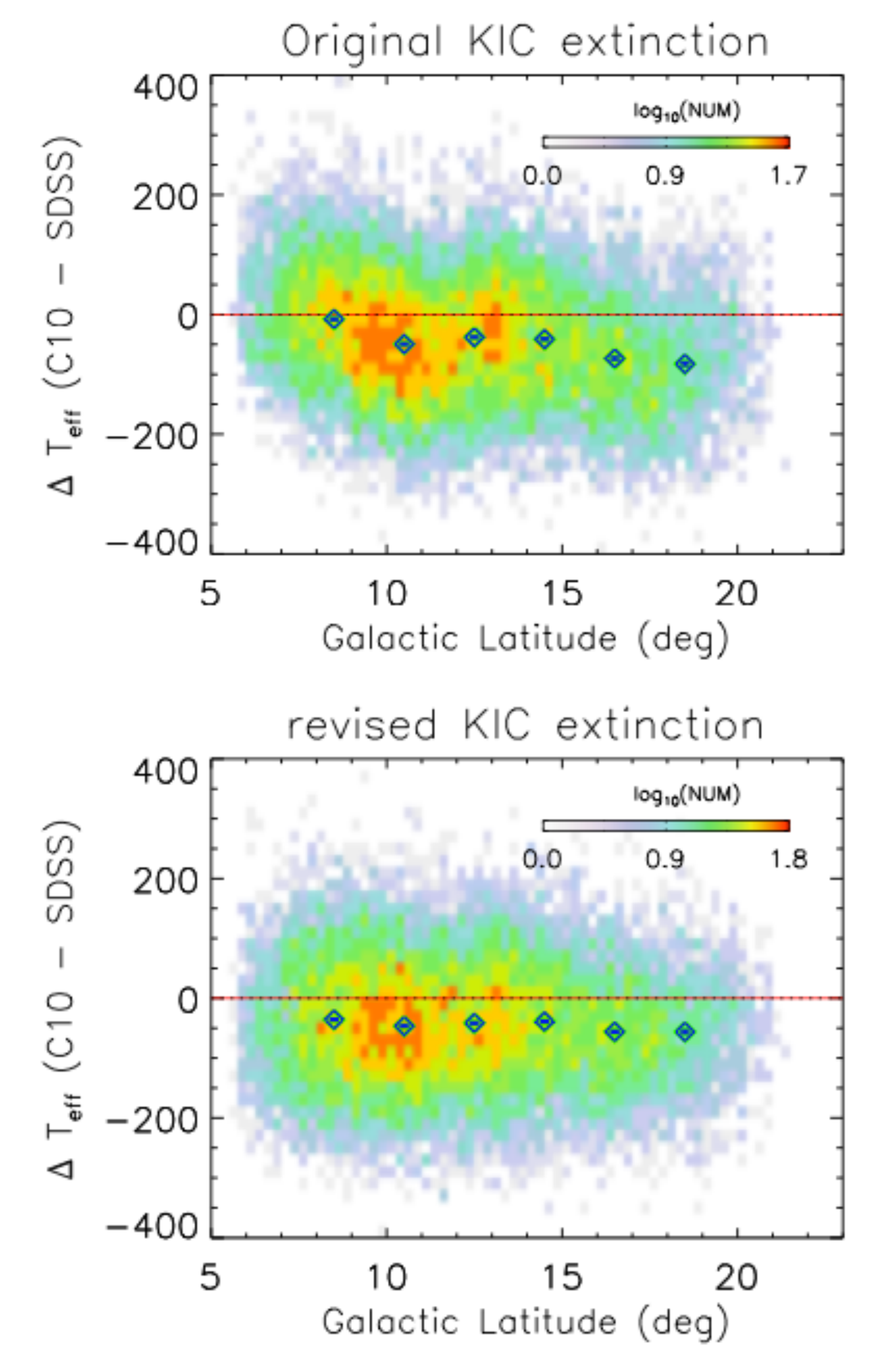}
  \caption{{\it Left:} Difference in $T_{\rm eff}$ from SDSS colors and from APOGEE/ASPCAP as functions of (top) Galactic longitude
    and (bottom) Galactic latitude, using the revised KIC reddening.  Compare to Figure~\ref{fig:sdss-aspcap-position}.
    {\it Right:} Difference in $T_{\rm eff}$ calculated with the IRFM method \citep{Casagrande_2010_irfm} and with SDSS {\it griz} colors, assuming
    (top) the original KIC reddening and (bottom) the revised KIC reddening.}
  \label{fig:sdss-aspcap-position2}
\end{center}
\end{figure}

In the left column of Figure~\ref{fig:sdss-aspcap-position2}, we show that the revised map clearly mitigates 
the spatial dependencies of the offsets between the photometric {\it griz} 
temperatures and the spectroscopic APOGEE ones.  In the top panel, the longitude structure is reduced 
(but not eliminated, suggesting further improvement can be made; \S\ref{sec:future}), and in the bottom panel, the latitude dependence has disappeared entirely.
Note that there is still a $\sim$200~K offset --- the revised reddening map has not solved the entire problem, but it does remove that
particular systematic contribution to the behavior.

The reddening revisions even appear to reduce some systematics in the comparison of photometric temperatures.
The right-hand column of Figure~\ref{fig:sdss-aspcap-position2} shows the difference 
between the IRFM and the {\it griz} $T_{\rm eff}$ values as a function of $b$, 
using the original KIC $E(B-V)$ (top) and the revised $E(B-V)$ (bottom).  
Both of these methods use extinction estimates but to different degrees.
Thus, an offset trend with $b$ is visible (though weaker than in the comparison to spectroscopic $T_{\rm eff}$)
but disappears with the revised reddening, most noticeably at higher $b$.

%\begin{figure} [!h]
%\begin{center}
%  \includegraphics[height=3.5in]{apokasc_LCirfm-vs-sdss-vs-latitude.pdf}
%  \caption{Difference in $T_{\rm eff}$ calculated with the Casagrande et al.\,2010 ``IRFM'' method and with the {\it griz} ``SDSS'' colors, assuming
%    (top) the original KIC reddening and (bottom) the rescaled KIC reddening.}
%  \label{fig:irfm-griz-lat}
%\end{center}
%\end{figure}

\section{Future Improvements} \label{sec:future}
\begin{figure}[!h]
%\begin{center}
\sidecaption[t]
  \includegraphics[trim=1in 1in 1in 1in, clip, width=6.4cm]{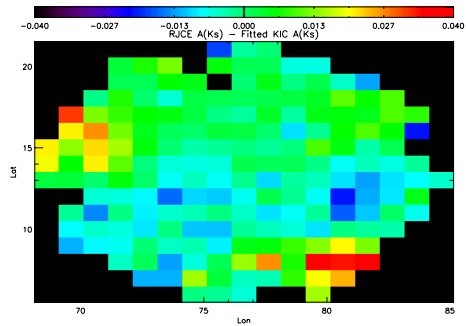}
  \caption{Difference between the revised KIC extinction map and the RJCE map, in $A(K_s)$.}
  \label{fig:residual}
%\end{center}
\end{figure}

A number of improvements to these preliminary results are actively in progress,
including incorporation of substructure parameterization into the KIC model fitting.
%\runinhead{Substructure}
The $\Delta E(B-V)$ residuals observed between the RJCE and the revised KIC maps are not dominated by random noise (Figure~\ref{fig:residual}).  
The presence of coherent clumps and gradients implies real structure in the ISM not accounted for by the KIC's exponential disk model.
For example, the residual map shows a clump at $(l,b) \sim (70.5^\circ,14^\circ)$, 
clearly visible in the RJCE and Schlegel maps but not in the smooth KIC one. 
This cloud corresponds to roughly 0.1 mag of $E(B-V)$ reddening {\it above} the background reproduced in the KIC map.
Thus, identifying and parameterizing these irregular variations will have a significant impact on many stars.

%A subtler feature is the presence of a ``tilt'' to the residual, with $\Delta E(B-V) \sim 0.1$ at $(l,b) \sim (80^\circ,8^\circ)$,
%a diagonal stripe of $\Delta E(B-V) \sim -0.02$ starting at lower longitudes, and nearly zero residual for $b > 13^\circ$.
%This behavior suggests the need for an additional longitudinal correction term, which is not surprising, since both the large-scale ISM density
%and the frequency of high density clumps depend on Galactocentric radius.

%\runinhead{Distances}
%\begin{itemize}
%\item Compare (i) KIC distances, (ii) APOGEE isochrone-based distances, (iii) KIC distances rescaled with seismic $R$ and either ASPCAP $T_{\rm eff}$ or {\it griz} $T_{\rm eff}$
%\item Redo distances from scratch, assuming seismic $R$, spectro/{\it griz} $T_{\rm eff}$, and revised KIC $E(B-V)$ 
%%(need to acquire BC$_\lambda$ and $m_{\lambda,\sun}$ from Brown et al.\,2011)
%\end{itemize}
%
%\runinhead{Extinction Law}
%\begin{itemize}
%\item Quantify impact of uncertainties in the assumed extinction law on the RJCE and revised KIC $E(B-V)$ values.
%\item Using the Stello et al.\,2013 evolutionary stages, identify the RC stars and measure CER$_\lambda(l,b)$ for $g < \lambda < 4.5\mu$
%\end{itemize}

\begin{acknowledgement}
%If you want to include acknowledgments of assistance and the like at the end of an individual chapter please use the \verb|acknowledgement| environment -- it will automatically render Springer's preferred layout.
GZ has been supported by an NSF Astronomy \& Astrophysics Postdoctoral Fellowship under Award No.\ AST-1203017,
and DA has been supported by the National Research Foundation of Korea to the Center for Galaxy Evolution Research (No.\ 2010-0027910).
\end{acknowledgement}

\vspace{-30pt}

%\input{referenc}
% BibTeX users please use
%\bibliographystyle{spphys}
%\bibliography{zasowski_sesto}

\begin{thebibliography}{14}
\expandafter\ifx\csname natexlab\endcsname\relax\def\natexlab#1{#1}\fi

\bibitem[{{Brown} {et~al.}(2011){Brown}, {Latham}, {Everett}, \&
  {Esquerdo}}]{Brown_2011_KIC}
{Brown}, T.~M., {Latham}, D.~W., {Everett}, M.~E., \& {Esquerdo}, G.~A. 2011,
  \aj, 142, 112

\bibitem[{{Casagrande} {et~al.}(2010){Casagrande}, {Ram{\'{\i}}rez},
  {Mel{\'e}ndez}, {Bessell}, \& {Asplund}}]{Casagrande_2010_irfm}
{Casagrande}, L., {Ram{\'{\i}}rez}, I., {Mel{\'e}ndez}, J., {Bessell}, M., \&
  {Asplund}, M. 2010, \aap, 512, A54

\bibitem[{{Drimmel} {et~al.}(2003){Drimmel}, {Cabrera-Lavers}, \&
  {L{\'o}pez-Corredoira}}]{Drimmel_2003_3dMWextinction}
{Drimmel}, R., {Cabrera-Lavers}, A., \& {L{\'o}pez-Corredoira}, M. 2003, \aap,
  409, 205

\bibitem[{{Drimmel} \& {Spergel}(2001)}]{Drimmel_2001_3dMWextinction}
{Drimmel}, R. \& {Spergel}, D.~N. 2001, \apj, 556, 181

\bibitem[{{Gonzalez} {et~al.}(2012){Gonzalez}, {Rejkuba}, {Zoccali}, {Valenti},
  {Minniti}, {Schultheis}, {Tobar}, \& {Chen}}]{Gonzalez_2012_bulgeextmap}
{Gonzalez}, O.~A., {Rejkuba}, M., {Zoccali}, M., {et~al.} 2012, \aap, 543, A13

\bibitem[{{Majewski}(2012)}]{Majewski_2012_apogee}
{Majewski}, S.~R. 2012, in American Astronomical Society Meeting Abstracts,
  Vol. 219, American Astronomical Society Meeting Abstracts \#219, \#205.06

\bibitem[{{Majewski} {et~al.}(2011){Majewski}, {Zasowski}, \&
  {Nidever}}]{Majewski_2011_rjce}
{Majewski}, S.~R., {Zasowski}, G., \& {Nidever}, D.~L. 2011, \apj, 739, 25

\bibitem[{{Marshall} {et~al.}(2006){Marshall}, {Robin}, {Reyl{\'e}},
  {Schultheis}, \& {Picaud}}]{Marshall_2006_3dMWextinction}
{Marshall}, D.~J., {Robin}, A.~C., {Reyl{\'e}}, C., {Schultheis}, M., \&
  {Picaud}, S. 2006, \aap, 453, 635

\bibitem[{{Molenda-{\.Z}akowicz} {et~al.}(2013){Molenda-{\.Z}akowicz}, {Sousa},
  {Frasca}, {Uytterhoeven}, {Briquet}, {Van Winckel}, {Drobek}, {Niemczura},
  {Lampens}, {Lykke}, {Bloemen}, {Gameiro}, {Jean}, {Volpi}, {Gorlova},
  {Mortier}, {Tsantaki}, \& {Raskin}}]{Molenda-Zakowicz_2013_keplerfollowup}
{Molenda-{\.Z}akowicz}, J., {Sousa}, S.~G., {Frasca}, A., {et~al.} 2013,
  \mnras, 434, 1422

\bibitem[{{Pinsonneault} {et~al.}(2012){Pinsonneault}, {An},
  {Molenda-{\.Z}akowicz}, {Chaplin}, {Metcalfe}, \&
  {Bruntt}}]{Pinsonneault_2012_temprevisions}
{Pinsonneault}, M.~H., {An}, D., {Molenda-{\.Z}akowicz}, J., {et~al.} 2012,
  \apjs, 199, 30

\bibitem[{{Pinsonneault} {et~al.}(2013){Pinsonneault}, {An},
  {Molenda-{\.Z}akowicz}, {Chaplin}, {Metcalfe}, \&
  {Bruntt}}]{Pinsonneault_2013_temprevisionserr}
{Pinsonneault}, M.~H., {An}, D., {Molenda-{\.Z}akowicz}, J., {et~al.} 2013,
  \apjs, 208, 12

\bibitem[{{Schlegel} {et~al.}(1998){Schlegel}, {Finkbeiner}, \&
  {Davis}}]{Schlegel_1998_dustmap}
{Schlegel}, D.~J., {Finkbeiner}, D.~P., \& {Davis}, M. 1998, \apj, 500, 525

\bibitem[{{Thygesen} {et~al.}(2012){Thygesen}, {Frandsen}, {Bruntt},
  {Kallinger}, {Andersen}, {Elsworth}, {Hekker}, {Karoff}, {Stello},
  {Brogaard}, {Burke}, {Caldwell}, \&
  {Christiansen}}]{Thygesen_2012_keplerfollowup}
{Thygesen}, A.~O., {Frandsen}, S., {Bruntt}, H., {et~al.} 2012, \aap, 543, A160

\bibitem[{{Zasowski} {et~al.}(2013){Zasowski}, {Johnson}, {Frinchaboy},
  {Majewski}, {Nidever}, {Rocha Pinto}, {Girardi}, {Andrews}, {Chojnowski},
  {Cudworth}, {Jackson}, {Munn}, {Skrutskie}, {Beaton}, {Blake}, {Covey},
  {Deshpande}, {Epstein}, {Fabbian}, {Fleming}, {Garcia Hernandez}, {Herrero},
  {Mahadevan}, {M{\'e}sz{\'a}ros}, {Schultheis}, {Sellgren}, {Terrien}, {van
  Saders}, {Allende Prieto}, {Bizyaev}, {Burton}, {Cunha}, {da Costa},
  {Hasselquist}, {Hearty}, {Holtzman}, {Garc{\'{\i}}a P{\'e}rez}, {Maia},
  {O'Connell}, {O'Donnell}, {Pinsonneault}, {Santiago}, {Schiavon}, {Shetrone},
  {Smith}, \& {Wilson}}]{Zasowski_2013_apogeetargeting}
{Zasowski}, G., {Johnson}, J.~A., {Frinchaboy}, P.~M., {et~al.} 2013, \aj, 146,
  81

\end{thebibliography}

%\section{References}
%References may be \textit{cited} in the text either by number (preferred) or by author/year.\footnote{Make sure that all references from the list are cited in the text. Those not cited should be moved to a separate \textit{Further Reading} section or chapter.} The reference list should ideally be \textit{sorted} in alphabetical order -- even if reference numbers are used for the their citation in the text. If there are several works by the same author, the following order should be used: 
%\begin{enumerate}
%\item all works by the author alone, ordered chronologically by year of publication
%\item all works by the author with a coauthor, ordered alphabetically by coauthor
%\item all works by the author with several coauthors, ordered chronologically by year of publication.
%\end{enumerate}
%You are encouraged to use BibTeX with the style \verb|spphys| provided with the template. Examples of citations are \cite{Kjeldsen95} and \cite{Reimers75}.

% Mac users: please ignore the error message: "! Package natbib Error: Bibliography not compatible with author-year citations."
\end{document}